\documentclass[sigconf,screen]{acmart}
\AtBeginDocument{%
  }

\setcopyright{acmlicensed}
\copyrightyear{2026}
\acmYear{2026}
\acmDOI{}

\acmConference[CHI '26]{ACM Conference on Human Factors in Computing Systems}{April 13--17, 2026}{Barcelona, Spain}

\usepackage{graphicx}
\usepackage{subcaption}
\usepackage{array}

\acmConference[PoliSim@CHI 2026]{PoliSim@CHI 2026: LLM Agent Simulation for Policy}{April 16, 2026}{Barcelona, Spain}
\renewcommand{\footnotetextcopyrightpermission}[1]{%
  \footnotetext{%
    This paper was prepared for \href{https://polisim.net/}{PoliSim@CHI 2026: LLM Agent Simulation for Policy}, a workshop at \href{https://chi2026.acm.org/}{CHI 2026 (CHI Conference on Human Factors in Computing Systems)}, April 16, 2026, Barcelona, Spain.
  }%
}

\begin{document}

\title{Human-Simulation Interaction}
\subtitle{From Prediction to Exploration in LLM Agent Simulations for Policy}

\author{Huanxing Chen}
\email{huanxing@stanford.edu}
\orcid{0009-0005-0895-3038}
\affiliation{%
  \institution{Stanford University}
  \city{Stanford}
  \state{California}
  \country{USA}
}

%% Article type: Research, Review, Discussion, Invited or position
\acmArticleType{Research}
%%
%% Links to code and data
%\acmCodeLink{https://github.com/borisveytsman/acmart}
%\acmDataLink{htps://zenodo.org/link}
%%
%% Authors' contribution
% \acmContributions{}
%%
%% Sometimes the addresses are too long to fit on the page.  In this
%% case uncomment the lines below and fill them accodingly.
%%
%% \authorsaddresses{Corresponding author: Ben Trovato,
%% \href{mailto:trovato@corporation.com}{trovato@corporation.com};
%% Institute for Clarity in Documentation, P.O. Box 1212, Dublin,
%% Ohio, USA, 43017-6221}
%%
%%
\begin{abstract}
Agent-based models have historically served as tools for generative explanation, constructing testbeds in which candidate micro-level behavioral rules can be tested for their capacity to produce observed macro-level phenomena. The integration of Large Language Models into agent-based simulation has expanded what these models can represent, but it has also introduced an unexamined shift in how users engage them. We argue that current generative agent-based models (GABMs) inherit the dominant interaction metaphor of conversational LLM interfaces - a question-answer pattern that positions users as consumers of system output rather than explorers of a possibility space. In the context of policy, where problems are wicked and ground truth is unknowable in advance, this metaphor produces a trust deficit that cannot be resolved through improved model accuracy alone. We open a design space we call human-simulation interaction, and argue that warranted trust requires interaction metaphors that restore the exploratory capacity simulation has historically supported.
\end{abstract}


\maketitle

\section{Introduction}
Agent-based modeling has long served social science as a tool for \textit{generative explanation} \cite{epstein2008model}. The classical workflow begins with an observed macro-level phenomenon (residential segregation, epidemic dynamics, market instability) and asks what micro-level rules, operating among heterogeneous agents in a defined environment, could produce it \cite{epstein1996growing, epstein2008model}. The modeler's task is not to predict but to isolate candidate mechanisms and construct a testbed in which to probe their sufficiency: if these simple rules are followed, does the macro pattern emerge? Epstein distinguishes this sharply from prediction - Plate tectonics explains earthquakes without predicting when the next one will strike. Yet as Epstein observes, the moment one posits a model, prediction - "as in a \textit{crystal ball} that can tell the future" - is reflexively presumed to be the goal \cite{epstein2008model}.

The recent integration of LLMs into agent-based simulations represents a significant expansion in what simulation can represent. LLM-powered agents can reason in natural language and operate in domains where the governing rules are unknown or too complex to formalize \cite{vezhnevets2023generative, park2023generative, larooij2025validation}, opening simulation to phenomena that previously resisted formal specification \cite{vezhnevets2023generative, zhang2026gplab}. But alongside this expansion in representational capacity, there has been a less examined shift in how users relate to the simulation itself. The locus of emphasis has drifted from generative explanation toward something more like prediction - from \textit{exploration} toward \textit{consultation}. This paper argues that this drift matters, and it matters especially for policy.

The trust deficit that surrounds LLM agent simulations for policy is widely recognized \cite{larooij2025validation, ghaffarzadegan2024generative}. The dominant response has been to improve simulation fidelity: validating that agent behaviors correspond to real human responses, calibrating agent architectures against empirical data, and demonstrating that emergent macro patterns track observed outcomes. Park et al. \cite{park2024generative}, for instance, construct agents from life story interviews with 1,000 individuals and evaluate whether those agents can predict the interviewees' survey responses - exemplifying a paradigm in which trust is pursued through representational accuracy. We do not dispute the importance of this work. But we argue that the trust question is also, and perhaps primarily, an interaction design problem. 

The issue is not that users consciously expect prophecy from simulation. A policymaker may approach a GABM seeking to update prior beliefs or develop intuition about a system's dynamics - not expecting definitive answers. But when the interaction takes the form of posing a scenario and receiving a narrative response, even these goals are channeled toward evaluating specific outputs rather than mapping the space of possibilities. A policymaker exploring how a housing regulation might affect displacement can, in principle, run varied scenarios - but when each interaction is linguistic and responsive, the natural rhythm becomes \textit{serial evaluation of individual outputs} rather than comparison across conditions. The understanding of which conditions matter and why, which only becomes visible through systematic variation, is what the consultation pattern fails to scaffold. When users engage a simulation through a crystal-ball metaphor, the question "can we trust this simulation?" becomes structurally unanswerable. No improvement in accuracy resolves a deficit that originates in the framing of the interaction itself.

We argue that warranted trust in LLM agent simulations for policy may require not only more accurate models but a fundamental rethinking of human-simulation interaction itself - the metaphors, affordances, and interaction patterns through which users engage simulations, and which shape what those users can think and do with them. The transition from ABMs to GABMs carried not only a gain in capability but an unexamined shift in interaction metaphor, and attending to this shift - restoring the user's sense of active exploration and control - is a necessary condition for making these simulations genuinely useful for policy.

\section{What is Simulation For?} 
The assumption that simulation's primary value lies in prediction is widespread but contested within the modeling community itself. Epstein \cite{epstein2008model} enumerates sixteen reasons other than prediction to build a model, among them explaining, illuminating core uncertainties, discovering new questions, training practitioners, bounding outcomes to plausible ranges, and disciplining policy dialogue. Bankes \cite{bankes1993exploratory} formalizes the distinction as one between two paradigms of model use. \textit{Consolidative} modeling is driven by what is known: it assembles established knowledge into a single credible model intended to serve as a surrogate for the target system. \textit{Exploratory} modeling assumes deep uncertainty, that there exist many plausible models rather than one correct one, and uses computational experiments to map how conclusions change across the space of plausible assumptions and model structures. Under exploratory modeling, the value of simulation lies not in the accuracy of any single run but in what traversing an ensemble of plausible runs reveals about the structure of the problem and the conditions under which strategies succeed or fail \cite{bankes1993exploratory, lempert2003shaping}.

This distinction is especially consequential for policy, where the problems simulation is asked to address are not merely complex but \textit{wicked}. Rittel and Webber \cite{rittel1973dilemmas}, writing explicitly about social policy, demonstrated that such problems resist definitive formulation - the problem definition itself is contested, evolving, and inseparable from the proposed solution. Buchanan \cite{buchanan1992wicked} extended this to design more broadly, arguing that productive design thinking operates within the fundamental indeterminacy of wicked problems rather than prematurely resolving that indeterminacy. Related traditions arrive at compatible insights: Schön \cite{schon2017reflective} argued that skilled practitioners construct the problems they solve through ongoing conversation with the situation; Dorst and Cross \cite{dorst2001creativity} showed empirically that expert designers work through simultaneous co-evolution of problem and solution spaces.

Together, these literatures suggest that simulation's deepest affordance for policy is not the production of answers but the support of \textit{exploration} - a mode of engagement in which the user's understanding of the problem, not just the solution, can shift. This is fundamentally different from evaluation, which presupposes a fixed problem definition and assesses candidate answers against it. If exploration requires the simultaneous evolution of problem and solution, then the cognitive demand is for interaction that supports both convergent and divergent cognition (in the sense of Guilford \cite{guilford1967nature}): generating alternative framings, surfacing unexpected dynamics, illuminating the structure of uncertainty - with the modeler able to move fluidly between modes, as Goldschmidt \cite{goldschmidt2016linkographic} has shown skilled designers do. The design question, then, is whether the way we interact with a simulation supports or suppresses this exploratory capacity.

\section{Metaphor Shift from ABM to GABM}
Traditional ABMs were natural vehicles for the kind of exploratory interaction that policy problems demand. Consider Schelling's \cite{schelling1971dynamic} segregation model: the modeler begins with an observed macro pattern (residential segregation) and hypothesizes a micro-level mechanism (mild same-type preference among neighbors). The model constructs an environment in which these rules operate, and the question is whether the macro pattern emerges. In this specific context, it does and overwhelmingly so. But beyond this, the modeler could vary the rules themselves, trying different preference thresholds, testing alternative mechanisms, forming and discarding hypotheses about which micro-level dynamics might matter, and in doing so treat the rules as objects of inquiry rather than fixed inputs. Schelling's significance was not predictive but explanatory: the model demonstrated that strong discriminatory preferences were \textit{not necessary} to produce segregation, and that mild preferences sufficed. In Epstein's \cite{epstein2008model} terms, it \textit{revealed the apparently simple to be complex}. And critically, the model's value was inseparable from the way users engaged it: by varying parameters, running batches, and comparing outcomes, users could map the conditions under which segregation did and did not emerge \cite{epstein1996growing, hammond2015considerations}. The interface affordances of ABM tools - parameter sliders, batch runs, outcome visualization - naturally supported Bankes' exploratory modeling methodology \cite{bankes1993exploratory}, even if not all practitioners used them that way.

Traditional ABMs thus supported exploration at two levels. At the macro level, users navigated a landscape of parameter configurations, running systematic experiments to understand how outcomes varied across assumptions. At the micro level, the rules themselves were transparent, legible, and open to revision: the modeler chose them as hypotheses about mechanisms, and their simplicity made the relationship between assumption and outcome traceable \cite{epstein2008model} - as Epstein and Axtell's Sugarscape \cite{epstein1996growing} classically demonstrated. ABMs served generative explanation \textit{through} exploratory interaction.

GABMs \cite{vezhnevets2023generative, park2023generative, larooij2025validation} gained something consequential in moving from rule-based to language-based agents. Where traditional ABMs required the modeler to hand-craft behavioral rules and specify utility functions, LLM-powered agents can draw on the common-sense reasoning and cultural knowledge embedded in their training to respond to situations in natural language, producing contextually sensitive decisions without explicit programming of each behavioral contingency \cite{vezhnevets2023generative, zhang2026gplab}. The Concordia framework \cite{vezhnevets2023generative} notes that GABMs "can incorporate far more of the complexity of real social situations" than rule-based models, and the GPLab system \cite{zhang2026gplab} demonstrates that LLM-based agents can overcome critical limitations of traditional ABMs in semantic understanding and adaptation to diverse policy contexts. These capabilities matter: they open policy simulation to domains that were previously inaccessible to formal modeling.

At the macro level, GABMs may initially appear to preserve the exploratory stance of traditional ABMs. The user can still configure scenarios, vary conditions, and compare outcomes. But on closer inspection, the micro level tells a different story. Each agent interaction now follows a question-answer pattern: the simulation poses situations to LLM-powered agents and receives natural language responses. We argue that this micro-level pattern tends to pervade the macro level of interaction as well. When individual agent behaviors look like question-answering, the entire simulation begins to feel like a system one consults for answers rather than a landscape one explores for insight. The parameter space (which was the primary object of interaction in ABMs) recedes behind a consultation pattern, and the user's cognitive stance shifts from navigator to evaluator. The modality of the parts sets the expectation for the whole: when every component interaction is linguistic and responsive, the system as a whole reads as something that speaks back rather than something one moves through. Even where the system exposes LLM reasoning traces, the resulting interaction is one of evaluation - assessing whether a particular agent responded plausibly - rather than exploration of the space of possible dynamics. This shift is most pronounced for downstream users - policymakers and decision-makers who encounter the simulation through its interface rather than its codebase.

This pervasion is not inevitable, but it is the default, and defaults shape thought. Research in ontological design suggests that designed artifacts do not merely serve functions but constitute the conditions of possibility for thought and action \cite{winograd1986understanding, willis2006ontological}. The issue is not that exploratory engagement with GABMs is impossible - users can run multiple scenarios, vary prompts, compare outputs - but that \textit{the inherited interaction metaphor does not invite or scaffold exploration}. Users must actively resist the gravitational pull of the crystal-ball frame. In Bankes' \cite{bankes1993exploratory} terms, GABMs operate in domains of deeper uncertainty than traditional ABMs (because the rules themselves are unspecified), yet the inherited interaction metaphor pushes users toward consolidative engagement: treating the simulation as a system that should converge on a credible answer rather than a space of possibilities to be mapped.

This consolidative pressure operates at the model level as well as the interaction level. LLMs are optimized for next-token prediction on high-probability outputs - a tendency that works against the generation of diverse, divergent scenarios \cite{zhou2025balancing}. The crystal-ball frame and the model's probability-maximizing tendencies are not independent problems but mutual reinforcements: the metaphor invites users to seek definitive answers, and the architecture is disposed to provide them.

When users are positioned as consumers of predictive accuracy in a domain that is definitionally unknowable \cite{rittel1973dilemmas}, the question "can we trust this simulation?" admits no satisfying answer. The interaction has framed an unanswerable question. LLM unreliability - hallucination, prompt sensitivity, cultural bias \cite{huang2025survey} - warrants serious technical attention, as does simulation-level fidelity \cite{larooij2025validation}. But the trust problem is partly independent of both: even a substantially more reliable GABM, interacted with through a crystal-ball frame, would still suppress the exploratory engagement that policy simulation most needs. Stadler et al. \cite{stadler2024cognitive} provide indirect support for this claim, demonstrating that Q\&A interaction with LLMs produces shallower reasoning even when system outputs are accurate. The cognitive cost is in the interaction pattern, not only in the output quality.

\section{Toward Exploratory Human-Simulation Interaction} 

If the inherited crystal-ball metaphor suppresses the exploratory engagement that policy simulation most needs, then the design question begins not with interface features but with metaphor: \textit{what should the user understand themselves to be doing with this simulation}? We emphasize the primacy of metaphor not because structural affordances are unimportant, but because users whose mental model of a system diverges from its design model will interact within the constraints of that model rather than the system's actual capabilities \cite{norman2013design, luger2016like}. A consultation model systematically suppresses exploratory use. The desktop metaphor did not merely add icons to command-line computing; it changed what users understood themselves to be doing when they used a computer in the first place - from issuing instruction to a complex machine to organizing materials in a workspace. Structural affordances (e.g. draggable icons, nested folders) followed from and reinforced the metaphor, but the metaphor was primary.

This framing suggests that the trust deficit surrounding GABMs for policy may be, in part, an unintended consequence of the inherited interaction metaphor. If users approach the simulation as a crystal ball, they will evaluate it on predictive accuracy - the one criterion it cannot satisfy in a wicked-problems domain. Reframing the interaction from prediction to exploration does not eliminate the need for reliable models, but it changes the terms on which trust is assessed. Trust in an exploratory tool is calibrated against process rather than output: not "did the system get the answer right?" but "did interacting with this system surface considerations I had not anticipated, and did varying conditions produce meaningfully different results?" This recalibration is not without its own risks - exploratory engagement can produce false confidence as easily as genuine insight. Emerging work on human-AI interaction suggests that alternative metaphors - AI as provocateur \cite{sarkar2024ai}, antagonist \cite{cai2024antagonistic}, coach \cite{hofman2023sports} - can sustain more active cognitive engagement than the default Q\&A pattern. The question is whether analogous reframings can work for simulation.

Rather than prescribing specific solutions, we raise several questions that we believe are central to this design space.

\subsubsection{\textbf{What is the unique affordance of LLM-based simulation, and what does it sacrifice?}} 
GABMs can represent domains that resist formalization - precisely the domains where policy operates. But in gaining expressiveness, they lose the transparent relationship between assumption and outcome that made traditional ABMs such effective tools for exploration. The challenge is not to replicate ABM-style parameter exploration in a GABM context, but to discover what forms of exploration are \textit{native} to language-based simulation. What does it mean to explore a space of prompts, personas, and narrative framings rather than a space of numerical parameters?

\subsubsection{\textbf{What attitude toward uncertainty should the interaction support?}}
Under a crystal-ball metaphor, divergence across simulation runs is noise - evidence of unreliability. Under an exploratory metaphor, divergence is signal, revealing the structure of the possibility space \cite{bankes1993exploratory, lempert2003shaping}. Yet communicating this productively remains a design challenge: Hullman \cite{hullman2019authors} documents a persistent gap between visualization authors' stated belief in the value of depicting uncertainty and the widespread norm of omitting it. Designing for exploration means designing for a relationship with uncertainty in which the user can hold multiple possibilities in consideration rather than rushing toward a single assessed answer - an attitude consistent with what Buchanan \cite{buchanan1992wicked} describes as operating within the fundamental indeterminacy of wicked problems rather than prematurely resolving it.

\subsubsection{\textbf{Who is the human in human-simulation interaction?}}
The modeler who constructs the simulation, the policymaker who consults it, and the populations whose behavior is modeled all stand in fundamentally different relationships to the crystal-ball problem. The modeler may never have operated within a consultation frame - their practice already involves hypothesis formation and iterative experimentation. The policymaker, by contrast, is the stakeholder most susceptible to the crystal-ball metaphor, and the one for whom the shift from prediction to exploration is most consequential. Meanwhile, Haghighi et al. \cite{haghighi2025ontologies} demonstrate that the ontological assumptions embedded in LLM agent architectures - what counts as memory, what constitutes a person, what makes an event significant - are neither neutral nor transparent, meaning that different stakeholders may not even share a common understanding of what the simulation represents. Whether reframing requires metaphorical shift, structural affordances, or both is likely stakeholder-dependent - and answering this is an empirical question central to the design space this paper opens.

\section{Conclusion}
LLM agent simulations offer genuinely new capabilities for policy, but these capabilities will remain underrealized if the interaction metaphor through which users engage them defaults to answer-seeking in a domain where no definitive answers exist. The persistent trust question surrounding GABMs for policy is in significant part an interaction design problem - a consequence of inheriting a metaphor mismatched to the wicked, indeterminate nature of policy problems.

This reframing speaks directly to the co-evolution of simulations and policy processes that the PoliSim workshop organizers call for. But co-evolution requires that the user's understanding of the problem can shift through interaction with the simulation - precisely what a crystal-ball metaphor forecloses. The co-evolution the workshop envisions requires as a precondition an interaction frame that supports it. If the crystal ball is the wrong metaphor, what is the right one? Simulation for policy, at its best, functions as a workshop - a space where the user is a participant constructing understanding, not a client awaiting answers, and where the purpose is illuminating uncertainties, discovering questions we did not know to ask, and disciplining our dialogue about what matters. Designing human-simulation interaction that supports this capacity is an HCI problem, and it deserves sustained attention from this community.
\newline

\begin{acks}
Thank you to Dr Jonne Kamphorst, my supervisor for my master's thesis, for providing valuable feedback for this paper as well as encouraging me to combine different threads of my interests across HCI and beyond. 
\end{acks}

\bibliographystyle{ACM-Reference-Format}
\bibliography{BIB}
\end{document}